\documentclass[]{aastex631}
\usepackage{graphicx}
\usepackage{subfigure}

\usepackage{ulem,xcolor,bm}
\definecolor{mygray}{gray}{0.7}
\definecolor{pblue}{cmyk}{0.6, 0.9, 0, 0.3}
\definecolor{dblue}{RGB}{0,0,139} 

\begin{document}

\title{The uncertainty in water mass fraction of wet planets}

\author[0000-0001-5451-3221]{Michael Lozovsky}
\affiliation{Astrophysics Research Center of the Open university (ARCO),\\
Department of Natural Sciences,\\
The Open University of Israel \\
4353701, Raanana, Israel}


\begin{abstract}
Planets with masses between Earth and Neptune often have radii that imply the presence of volatiles, suggesting that water may be abundant in their interiors.
However, directly observing the precise water mass fraction and water distribution remains unfeasible. In our study, we employ an internal structure code MAGRATHEA to model planets with high water content and explore potential interior distributions. Departing from traditional assumptions of a layered structure, we determine water and rock distribution based on water-rock miscibility criteria. We model {wet planets} with an iron core and a homogeneous mixture of rock and water above it. At the outer regions of the planet, the pressure and temperature are below the rock-water miscibility point (the second critical point), causing the segregation of water and rock. Consequently, a shell of water is formed in the outermost layers. By considering the water-rock miscibility and the vapor state of water, our approach highlights the uncertainty in estimating the water mass fraction of detected exoplanets.

\end{abstract}


\section{Introduction} \label{sec:intro}

{Water-rich planets with masses between Earth and Neptune represent a diverse and largely mysterious class of worlds. }
These celestial bodies present a unique challenge due to the limited knowledge about their interior composition and structure, which is essential for unraveling the mysteries surrounding their formation and evolution.

These planets form through the accumulation of solids and gases within protoplanetary disks, acquiring distinctive characteristics influenced by their specific formation locations. Originating beyond the water ice line \citep{Lodders2003, Marcus2010,Mousis2020b}, these planets contain a significant water presence — either evident on the surface above the rocky layer or intricately mixed within the rock \citep[e.g.][]{Dorn2021, Kempton2023, luo2024, Shorttle2024}. A radius gap distinguishing between wet planets and dry planets was found by several groups\citep{Fulton2017, Burn2024, Parc2024, Schulze2024, Venturini2024}. The studies revealed a density distinction between the two populations, highlighting the significant role of atmospheres and evaporation. 


Traditionally it was assumed that as planets accumulate material from protoplanetary disks, they undergo differentiation, developing distinct layers based on the chemical separation of components \citep[e.g.][]{Stevenson2013,Madhusudhan2020,Nixon2021,Lozovsky2022}. This gives rise to a layered structure with an inner iron core surrounded by rock, and an atmosphere primarily composed of gases, including water steam. However, \citet{Vazan2022} challenged this conventional view, highlighting experiments  \citep[e.g.,][]{Grove2006,Melekhova2007,Kessel2015,Kim2021} {that demonstrate} the miscibility of rock and water under high pressure (P) and temperature (T) conditions, that are expected below the surface.

Depending on the temperature and pressure of the outermost layers, a water-rich rocky planet may undergo sublimation, leading to the formation of a primordial water vapor atmosphere. This initial water vapor atmosphere could experience partial mass loss{ through various mechanisms} \citep[e.g.,][]{Lammer2003, Owen2017, Howe2020, Aguichine2021, Pierrehumbert2023}, including impacts by planetesimals \citep{Schlichting2015, Lozovsky2023}, {and magma-envelope chemistry \citep[e.g.][]{Kite2020,Kite2021}}. Consequently, the observed planets may not accurately reflect the initial chemical composition of their original formation environment. {However, in this work we do not model the evolution of atmospheric and interior composition, but rather model a static structure for planets with a wide range of}
of water fractions (discussed in Section \ref{subsec:compos}) to account for the broad range of possible evolutionary outcomes.

In our study, we delve into the relationship between the water and rock mixture within planetary interiors, and the mass of a planet's water steam atmosphere. Specifically, we concentrate on the interplay between rock, represented  by SiO$_2$, and water (H$_2$O) in planets up to 10 M$_\oplus$. 
We draw upon the research of \citet{Vazan2022}, which outlined the conditions under which separation or mixing of rock and water occurs: with pressure-temperature conditions above second critical point (SCP) the components are expected to be in homogeneous mix, while below this point they are expected to segregate (for a detailed exploration of various pressure-temperature (P-T) {separation} conditions, refer to the Appendix 1). 
Under these physical circumstances, the outer layers may separate from the rock-water mixture, resulting in the formation of an outer water layer. 
{If temperature and pressure in this layer are high enough, the atmosphere is in a vapor state, i.e., water-steam atmosphere.}
The layers beneath maintain conditions surpassing SCP, persisting in a mixed supercritical fluid state. This configuration paints a picture where the outermost water layer remains relatively shallow compared to the older layered models, with a pure rocky mantle positioned beneath a pure water layer. {In more realistic scenarios, } planetary {mantles} are expected to exhibit a compositional gradient, where materials become progressively heavier with increasing depth \citep[e.g.,][]{Unterborn2020,Dorn2021}. Since the exact properties of the gradient depends on many unknown parameters, in our work, instead of studying possible gradual structures, we compare the two end-members of homogeneous mixture versus full rock and water segregation.



While water-rich planets originally formed beyond the water ice line, various models propose that inward planetary migration within the protoplanetary disk could bring these objects closer to their host stars \citep[e.g.,][]{D_Angelo_2016, Raymond2018, Izidoro2022}. Our investigation focuses on planets with temperatures ranging from 400 K to 1000 K, {in which many observed of planets of sub-Neptune mass are detected.}
{Planets with equilibrium temperatures above 300K are predicted to be in a post-runaway greenhouse stage \citep[e.g.][]{Kopparapu2013,Turbet2019}. We do not consider pre-runaway greenhouse planets in this study, and use $T_{eq}$=400 K as the lower boundary for our models (models with $T_{eq}$=300 K are shown in the Appendix). As vapor atmospheres are central to our study, we also exclude very high temperatures (above 1000K) to prevent significant water mass loss due to photoevaporation \citep[e.g.][]{Kurosaki2014}. While the temperature effect on the atmospheric radius is significant, its effect on the deep interior thermal evolution shown to be relatively small compared to other effects \citep[][]{Vazan2022}. }

Whether considering the older model with complete separation of water and rock or our new model, where most water is trapped within the rock beneath a layer of pure water - both yield pure water in the upper atmosphere. However, fully differentiated models generally result in higher radii for the same mass, bulk global composition, and equilibrium temperature compared to mixed models \citep[e.g.][]{Dorn2021,luo2024}. This is due to a large mass fraction of the lighter component (water) located in the lower-pressure outer layers.



In this study, instead of imposing a predetermined vapor atmosphere mass, we enable the mixture to undergo separation based on the P-T conditions outlined in \citet{Vazan2022} and \citet{Melekhova2007}. Primordially, we consider the mixture to be a fully blended homogeneous combination of rock and water. Then, the outer planetary region where the separation occurs is then modeled as pure water.


\section{Methods}

In this framework, we apply the MAGRATHEA internal structure code, developed by \citet{Huang2022}. This open-source planetary structure code is tailored for fully differentiated spherically symmetric interiors, solving standard planetary structure equations. It outputs internal structure model and basic planetary parameters, like radius (R), pressure (P), temperature (T), and density ($\rho$) as functions of mass (M). Differing from the original framework of MAGRATHEA of employing separate equations of state (EOS) for water and rock to different regions of the planet, we adopt a single EOS for the mixture for regions in which the miscibility criterion is fulfilled.

\subsection{Planetary Composition Models}

\begin{figure}[h]
    \centering

    \includegraphics[width=0.65\textwidth]{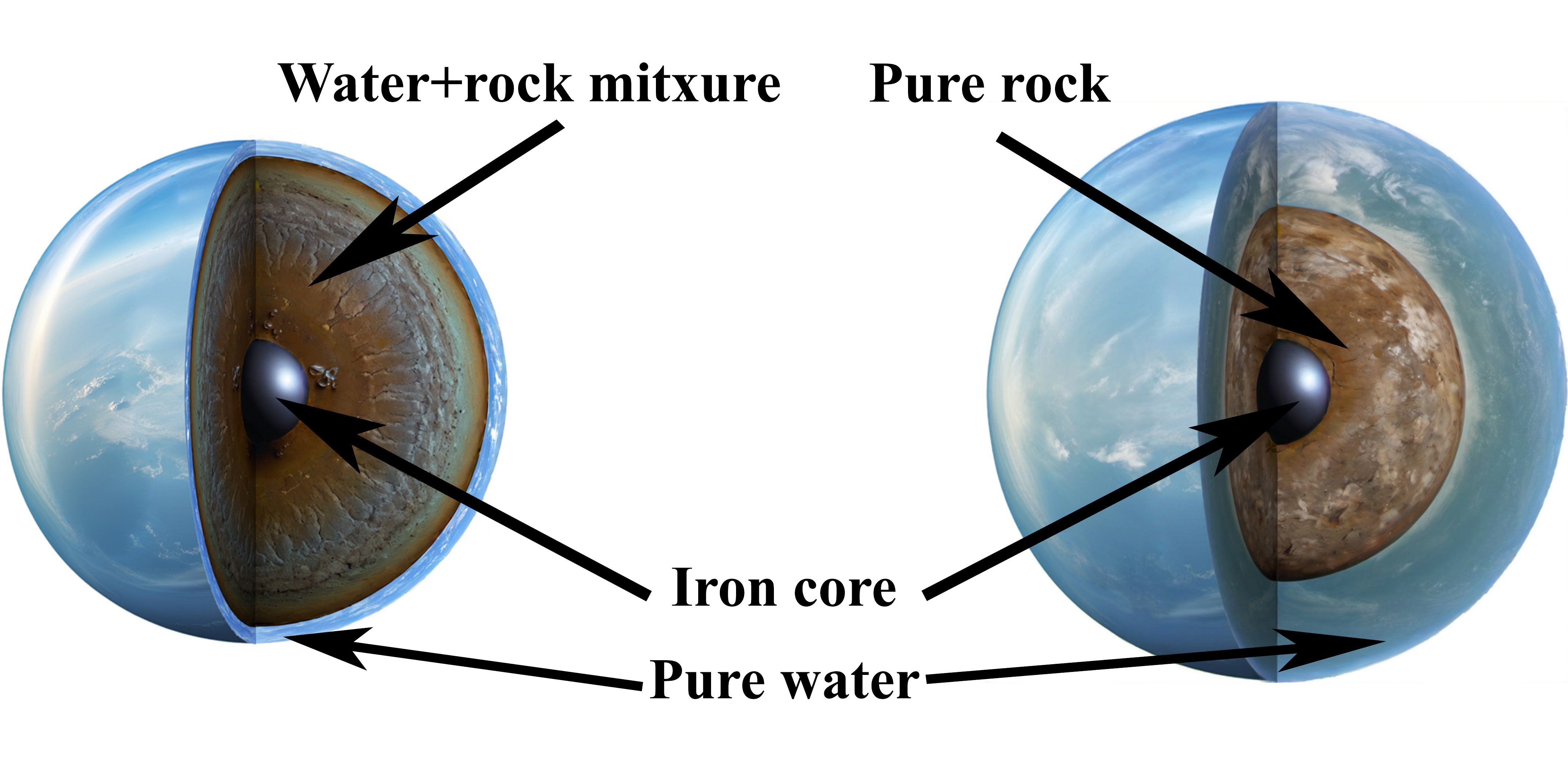}
    \caption{Planets composed of iron, rock, and water. Illustrations not to scale. The planet model on the left-hand side develops a shallow layer of pure water, while the majority of the water is locked in the rock {("mixed")}. The model on the right-hand side is assumed to have a three-layer structure {("separated")}. The physical properties of the structures are shown in Figure \ref{fig:RhoR5M500K}. }\label{fig:Illustration}
  
\end{figure}

Our study centers on planets with masses ranging from 1 to 10 Earth masses ($M_\oplus$), predominantly composed of rock (SiO$_2$), water (H$_2$O), and iron (Fe). We posit that these planets originated beyond the water ice line in protoplanetary disks and subsequently migrated inward, while maintaining high water mass fractions (WMF) (up to 0.5). 
The planets are assumed to be fully formed, in hydrostatic equilibrium, and spherically symmetric.
{The outer radius of a planet is defined at the 100 mbar pressure level, following \citet{Huang2022}. This choice reflects a compromise between observational relevance and model consistency, as 100 mbar is commonly associated with the pressure level probed in optical transits of clear atmospheres} \citep{Grimm2018}, {and is within the physically plausible range of transiting pressures used in atmospheric models(1--100 mbar) \citep{Aguchine2024}.} \footnote{{Some studies use a surface pressure of 1 mbar or even 1~$\mu$bar to match high-altitude haze formation layers \citep[e.g.][]{Turbet2019, Tang2024, Aguchine2024}, this choice depends strongly on wavelength and cloud treatment, and  shifts the radius upward.}}

We study two end member configurations: fully separated rock and water model versus a model that includes the miscibility of these two components. In both of the scenarios, we assume the innermost layers to be an iron core that is fully segregated from the mantle. In both scenarios, there is a homogeneous water layer at higher altitudes, which {is expected} be in the vapor phase, as elaborated in the subsequent sections. The two end members cases  are later referred as ”mixed” and ”separated", see Figure \ref{fig:Illustration}.



\subsubsection{Numerical Simulations and EOS}
\label{sec:EOSmodel}

To model the structures of these planets {in both "separated" and "mixed" scenarios we apply MAGRATHEA code \citet{Huang2022}}. MAGRATHEA operates under the assumption of fully differentiated spherically symmetric interiors, with the essential input parameters being the surface temperature and chemical composition of each layer, along with the mass of each layer. The code iteratively solves the boundary conditions of the hydrostatic equations to determine the planetary structure. The temperature and pressure at the boundaries of different planetary layers are conserved (One can find a detailed explanation of the code and the solved structure equations in \citet{Huang2022}).

A limitation of this software lies in its assumption of sharp transitions between materials and its inability to handle mixtures. In the absence of an established EOS for rock-water mixtures, we derive one by combining two well-established EOS models: AQUA for water \citep{Haldemann2020} and FEOS for rock \citep{Faik2018} (which is an improved version of QEOS by \citet{More1988}  and  a development of the MPQeos code by \citet{Kemp1998} ). Both EOS models are publicly available, covering extensive pressure, temperature, and density (P-T-$\rho$) ranges while accounting {different phases and transitions between them}
: {AQUA spans pressures from $10^{-7}$ to $10^5$ bar and temperatures from $10^2$ to $3\cdot 10^4$ K, while FEOS covers densities from $10^{-6}$ to $10^4$ g/cm$^3$ and temperatures up to $10^9$ K.}



The water-to-rock ratio in the mixture serves as one of our free parameters in this study, set separately for each run. We calculate the thermodynamic properties of the water-rock mixture by calling the EOS of the two compounds individually. Then,  instead of pre-creating a single table of mixed material properties, {our code uses a modified version of MAGRATHEA \footnote{As we explaiend, the original MAGRATHEA was designed for single-material-layers calculations, whereas our implementation dynamically combines EOS contributions during runtime.} that treats the water-to-rock ratio as an input parameter and calculates the properties of the mixture on the during the run}. The density of the mixture for a given composition, T and P is calculated as: 


\begin{equation}
\frac{1}{\rho_{\text{mix}}} = \frac{X_w}{\rho_w} + \frac{X_r}{\rho_r},
\end{equation}

where $\rho_{\text{mix}}$ is the density of the mixture, and $X_w$ and $X_r$ are mass fractions of water and rock in the mixture, with $X_w + X_r = 1$. $\rho_w (P,T)$ and $\rho_r (P,T)$ are densities of water and rock, calculated by interpolating from AQUA and FEOS, respectively, for given P and T.

The MAGRATHEA code requires a $\frac{dT}{dP}$ profile. We assume an adiabatic profile, so the temperature profile of the mixed material is calculated at constant entropy as {(see Appendix 2)}:

\begin{equation}
\left( \frac{dT}{dP} \right)_{s,\text{mix}} = -\frac{X_w \left(\frac{\partial s}{\partial P}\right)_{T, w} + X_r \left(\frac{\partial s}{\partial P}\right)_{T, r}}{X_w \left(\frac{\partial s}{\partial T}\right)_{P, w} + X_r \left(\frac{\partial s}{\partial T}\right)_{P, r}}.\label{dTdPeq2}
\end{equation}

where $\left( \frac{dT}{dP} \right)_{s,\text{mix}}$ is adiabatic profile of a mixture . The partial derivatives are calculated as:

\begin{eqnarray*}
\left( \frac{\partial s}{\partial P} \right)_{T, r} & = & \frac{S_{r}(P + \Delta P, T) - S_{r}(P - \Delta P, T)}{2 \Delta P} \\
\left(\frac{\partial s}{\partial P}\right)_{T, w} & = & \frac{S_{w}(P + \Delta P, T) - S_{w}(P - \Delta P, T)}{2 \Delta P} \\
\left(\frac{\partial s}{\partial T}\right)_{T, r} & = & \frac{S_{r}(P, T + \Delta T) - S_{r}(P, T - \Delta T)}{2 \Delta T} \\
\left(\frac{\partial s}{\partial T}\right)_{P, w} & = & \frac{S_{w}(P, T + \Delta T) - S_{w}(P, T - \Delta T)}{2 \Delta T},
\end{eqnarray*}

where $S_w$ and $S_r$ are entropies of water and rock calculated by direct interpolations from the AQUA and FEOS, respectively. $\Delta T$ and $\Delta P$ are small differentials, taken as 0.1\% of the corresponding T and P.

{Due to numerical discontinuities between adjacent cells in the AQUA EoS, which we use to simulate water, our gradient calculation method can suffer from reduced accuracy near these boundaries. While AQUA smooths sharp physical phase transitions, derivatives computed across adjacent cells (e.g., at points \(P,T\) and \(P,T+\Delta T\) in different cells) may introduce errors due to the bilinear interpolation’s inherent cell-edge discontinuities. These errors can accumulate along the adiabat when iteratively calculating derivatives. The estimation of the impact of this effect is discussed in section \ref{sec:Discussion}.}  

The innermost part of the planet (the core) is modeled as pure iron (Fe). The iron phase transitions are accounted for, as the EOS of iron and the melting curves are taken from \citet{Smith2018}, \citet{Dorogokupets2017}, and \citet{Bouchet2013}.
{The temperature profile does not include discontinuities between layers of different compositions.}
Also, we neglect the entropy of mixing when calculating the adiabatic \(\frac{dT}{dP}\) profile, as it is assumed to be a small value compared to the entropy of water and rock \citep[e.g.][]{Pan2023}. 

Among the most important outputs of the code is the outer planetary radius for a given combination of total mass, assumed composition and structure and equilibrium temperature ($T_{\text{eq}}$), as this {radius} can be directly compared with observations.

\subsubsection{Separation Processes}
\label{sec:mixture-separation}



In "separated" scenarios, we assume a priori that there is full segregation between water and rock; Conversely, in "mixed" scenarios, we apply a criterion for water rock miscibility. When the temperature and pressure in the outer layers fall below  SCP values, the rock sinks down to the mixed layers below, so all layers above the critical point are considered to be rock-free. These water layers are modeled using $X_r=0$ (and therefore pure AQUA EOS \citep{Haldemann2020}), while the layer below are considered to be homogeneous mix (see section \ref{sec:EOSmodel}). {AQUA EOS determines the phase of the water in the outer shell, including vapor, {supper-critical state} or solid ice, depending on P and T. As the upper layers are now modeled as pure water, the rock mass that was in these layers is added to the layers below the pure water layer, that are remain rock and water mixture. This is done by gradually increasing rock:water ratio of the mixed region, until reaching the total rock:water ratio equal to the expected one, with error up to  1\%. }
In $T_{eq}$ above 500K the effect of this access rock on water:rock:iron ratio of the planet is less than 0.1\%, which is below our precision.


SCP, that is considered to be the critical conditions for water to separate from the rock-water mixture is taken to be log(T) = 3.3 (in K) and log(P) = 1.1 (in GPa), following \citet{Melekhova2007}. Alternative SCP's are discussed in the Appendix 1. It should be noted that the true transition from full miscibility to imiscibility is expected to be gradual over a small region and not sharp. The size of this shallow transition region is expected to relatively small, compared to the region of pure rock  \citep{Vazan2022}, 
{and modeled as pure water in this study.}
\footnote{{Our modeled planets have about 1e-5 of their total mass in the transition region. See section \ref{src:limits}.}}

\subsubsection{Composition}
\label{subsec:compos}


The MAGRATHEA code employs various parameters to simulate planetary structures. The critical ones are total planetary mass, outer layer temperature (that is taken to be equilibrium temperature $T_{\text{eq}}$), and composition, namely mass ratios of iron, rock, and water. In addition, we add one more crucial parameter of material distribution ("mixed" versus "separated)". 

In this study we focus on planets up to 10 M$_\oplus$
composed of rock, water and iron only. The iron is always located in the center, forming a core. {The thermal profile of the iron core is chosen to be adiabatic; (See section 3.3.1 of \citet{Huang2022} for details )}. This core may have its own internal structure with liquid and solid parts, following EOS of iron (see section \ref{sec:EOSmodel}), but does not include any other chemical compounds. As we assume a distinct core-mantle boundary, water may exist solely within the rock, without considering the possibility of an iron-water mixture \citep[][]{luo2024}. 
As our models are water-rich, the contribution of the outer water shell on the total radius is significantly larger compared to the contribution of the core (Figure \ref{fig:Hist5Me}).





{We assume that the ratio of rock to iron is 2:1 by total mass, which is considered as a rough approximation to terrestrial composition \citep[e.g][]{Mordasini2015} .}
Therefore, if the water mass fraction is chosen to be $W$, then the iron mass fraction is calculated as $I = \frac{{1 - W}}{3}$, and the rock mass fraction is calculated as $R = 2I$. This ratio scheme holds both for "separated" and "mixed" scenarios, as separation does not change the bulk composition. { As we focus on water-rich planets, {WMF} for the comparative analysis varies from {0.05} to 0.5. We chose {0.05} as a lower bound in order for the water distribution effect to be clearly seen. The upper bound of {0.5 was selected based on indications from disk properties} \citep[e.g][]{Lodders2003, Zeng2019}. Later, when modeling specific exoplanets (section \ref{sec:exopl}), we do not limit the the WMF, and model planets with much smaller WMF {(less then 0.01)}. }

\subsection{Finding WMF for given planets}
\label{sec:findWMF}

{In Section \ref{sec:exopl}, we use MAGRATHEA to calculate the WMF for a group of discovered exoplanets. As the code takes WMF as an input rather than providing it as an output, we invert this relationship by employing an adaptive, damping-controlled root-finding algorithm that iteratively adjusts the WMF until the modeled planetary radius matches the observed value within 0.01 R$_\oplus$\footnote{0.01 R$_\oplus$ is the typical precision of the exoplanet databases}. This procedure, a variation of the bisection method, efficiently handles the strong sensitivity of the planetary radius to small changes in WMF.}
{M, R, and T are set to the central observed values, without considering the measurement uncertainties. This procedure was done both for "mixed" and "separated" models.}

\section{Results}

\begin{figure}[h]
    \centering
   \includegraphics[width=1.0\textwidth,trim={1.9cm 1.5cm 1.9cm 0.8cm},clip]{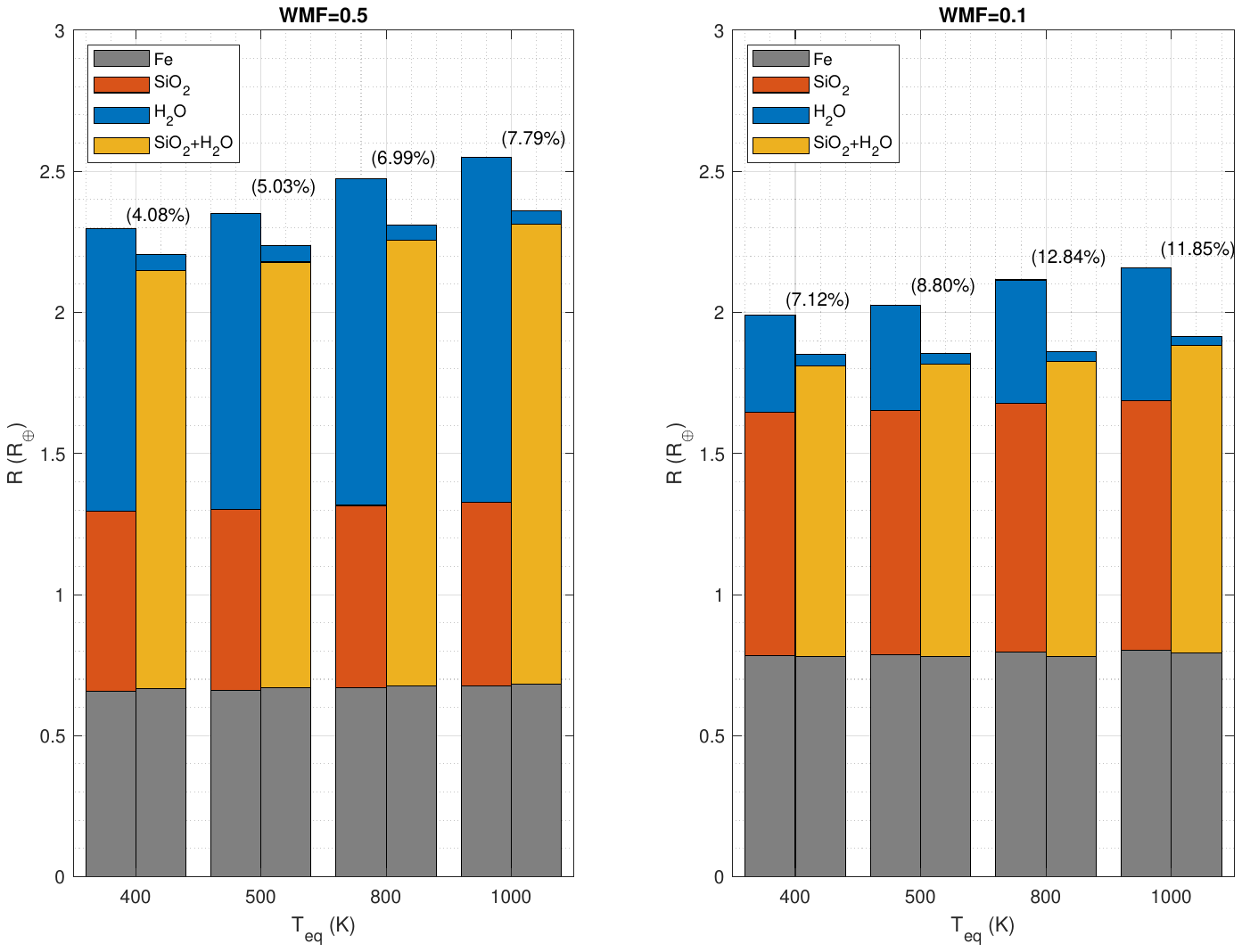}
    \caption{Representation of the internal structure of planets with a mass of 5M$_\oplus$ and a WMF of 0.5 {(left panel) and 0.1 (right panel)}. The left columns depict planets with a sharp separation between rock and water, while the right columns represent models featuring fully mixed rock/water interiors.  The percents above each pair of columns show the relative difference in the planetary radius {between "separated" and "mixed"}. In the right-column models {("mixed")}, a pure water layer develops on top of the planets, following the separation of water from rock under the pressure and temperature conditions (SCP) described by \citet{Melekhova2007}. Note some non-{monotonic}  behavior of the differences in high temperature and WMF of 0.1: this is a result of larger fraction of water mass mixed in the rock. An example of detailed interiors are shown in Figure \ref{fig:RhoR5M500K}.} 
    \label{fig:Hist5Me}
\end{figure}

\begin{figure}[h]
    \centering
    \includegraphics[width=1.0\textwidth]{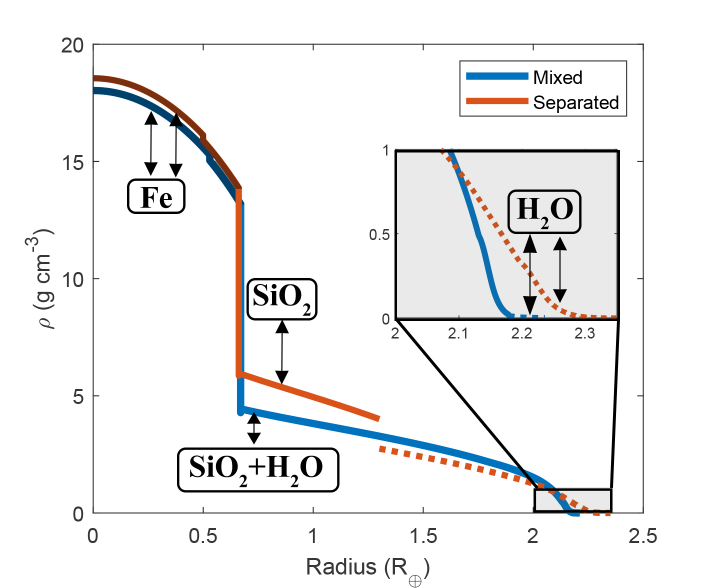}
    \caption{{Internal structure of planets with 5 M$_\oplus$ and equilibrium temperature of 500K, with separated and mixed mantles. Both of the models have the same bulk composition of  WMF=0.5 (2.5 M$_\oplus$ of water, 1.67 M$_\oplus$ of rock and 0.83 M$_\oplus$ of iron). The darker colors mark the iron core, and the dashed lines indicate outer water shells. Note the jumps in the density: {jumps inside the iron core} correspond the phase transitions \citep{Smith2018,Dorogokupets2017} .{ The jump from solid lines to dotted lines indicates transition from regions of the mantle to pure water layers. }} }
    \label{fig:RhoR5M500K}
\end{figure}

{The internal structures of planets with a mass of 5 $M_\oplus$, WMF of 0.1 and 0.5 and temperatures between {400 K} and 1000 K are  presented in figure \ref{fig:Hist5Me}. For each T and WMF, we compare the interiors of "separated" (fully differentiated) models to "mixed" (models where the outer shell is water that separated form the mix; see Section \ref{sec:mixture-separation}).  {At 400 K}, separated and mixed cases appear to have similar external radii (up to {7.12\%} difference) for both WMF, while having very different internal structures. In the both "separated" and "mixed" cases the external radius is increases with the temperature, and also the differences between "separated" and "mixed" are increasing with the temperature. The "mixed" cases in both panels demonstrate that the outer pure water shell becomes shallower as temperature rises, {because} the critical T and P (SCP) are reached closer the the surface. Thus, as less water is present on the surface, more water incorporated within the rock-water mix {(this point is deliberated later in section \ref{sec:MR_rel} and shown in in Figure \ref{fig:upper_part})}. Also, temperature governs other process: {thermal expansion of the vapor on the outer shells, that contributes to increasing the total radius.}
\\ As expected, at higher temperatures, more notable distinctions emerge between "mixed" and "separated" scenarios, with the fully separated scenario exhibiting a significantly larger radius for equivalent temperature and bulk composition. The reason for this greater difference at higher temperatures is that the water layer becomes shallower, as the location of the  SCP is closer to the surface. The differences between "mixed" and "separated" cases get larger with smaller WMF, as there is less water { available as vapor} in the "mixed" case. \\Figure \ref{fig:RhoR5M500K} shows an example of internal structures of planets with equilibrium temperature of 500K and mass of 5$M_\oplus${, highlighting the difference between the two scenarios.}}

\subsection{Exoplanets}
\label{sec:exopl}

\begin{figure}[h!]
    \centering
    \includegraphics[width=0.9\textwidth, trim={1.2cm 6.5cm 1.2cm 7.0cm},clip]{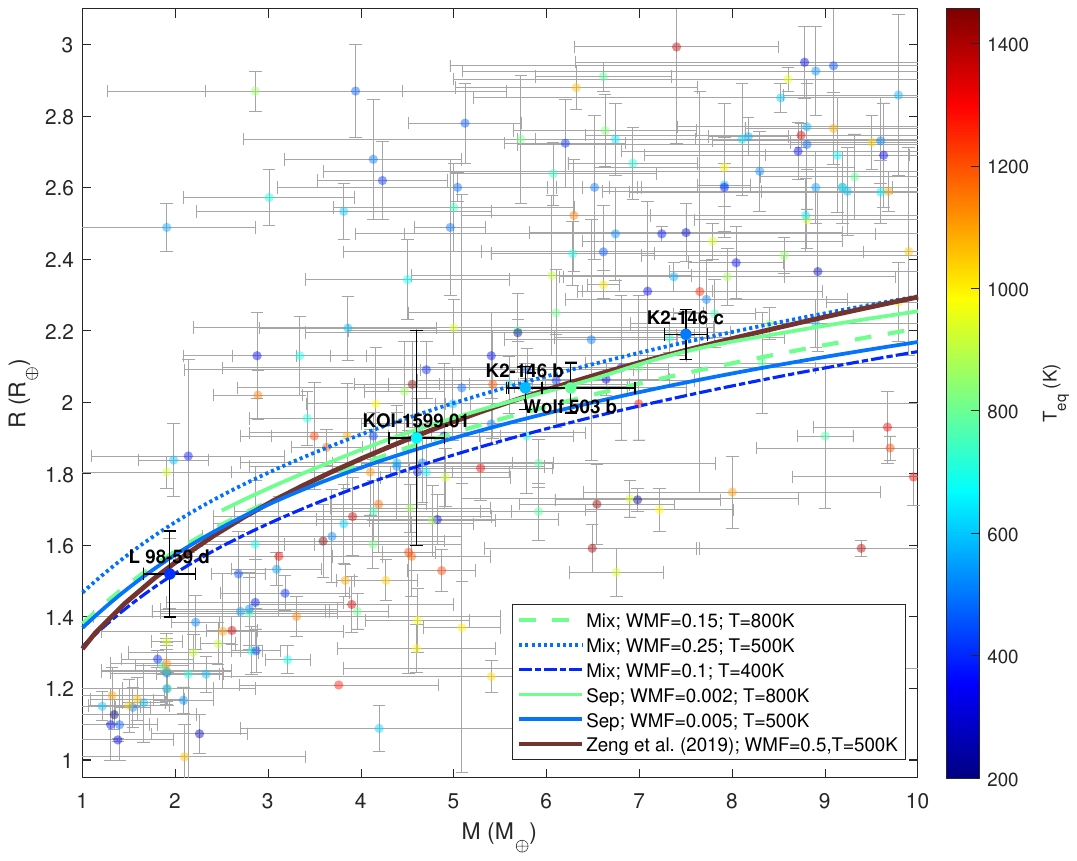}
        \caption{{Mass-Radius-Temperature diagram showing a sample of detected exoplanets with the selection of exoplanets highlighted. The selected sub-subsample can be identified as having WMF close to 0.5, based on M-R relations form \citet{Zeng2019} (brown solid curve). The dashed, dotted and dash-dotted curves are to show typical M-R relation for "mixed" scenario for different WMF and T, the solid curves are for "separated" scenario. For wider picture of M-R relations see figure \ref{fig:MRTWMF300_500}.}} The exoplanet sample is based on \textit{exoplanet.eu} publicly available catalogue, limited to planets between 1$M_\oplus$ and 10 $M_\oplus$. 
    \label{fig:MvrRexo}
\end{figure}

{Figure \ref{fig:MvrRexo} presents a sample of discovered exoplanets with developed M-R relations. As shown in the figure,}  the majority of  planets lying in a mass regime of 2-10 $M_\oplus$  have a relative high radius and thus {expected to have} significant mass fraction of volatiles.
\citet{Luque2022} demonstrated that among M-dwarf stars the mean planetary density distinguishes between two groups of sub-Neptune-sized planets: rocky worlds and water worlds. {Using mass and radius measured values,} they identified a subset of planets 
that align well with a theoretical mass-radius curve corresponding to a planet with a 0.5 water mass fraction. {However, subsequent studies (e.g., \citet{Burt2024, Fridlund2024,Dimitris2025}) have revised the mass and radius estimates for some of these planets, which would alter their inferred compositions. Also, it should be noted that the 0.5 water line \citet{Luque2022} used in their study} is derived from an analytical approach assuming a solid state layered structure, without considering temperature effects and vapor atmosphere{(the M-R curve based on \citet{Zeng2019})}. 


\begin{figure}[h!]
    \centering
    \subfigure{
        \includegraphics[width=0.54\linewidth, trim={0.5cm 10.5cm 0.0cm 11.0cm},clip]{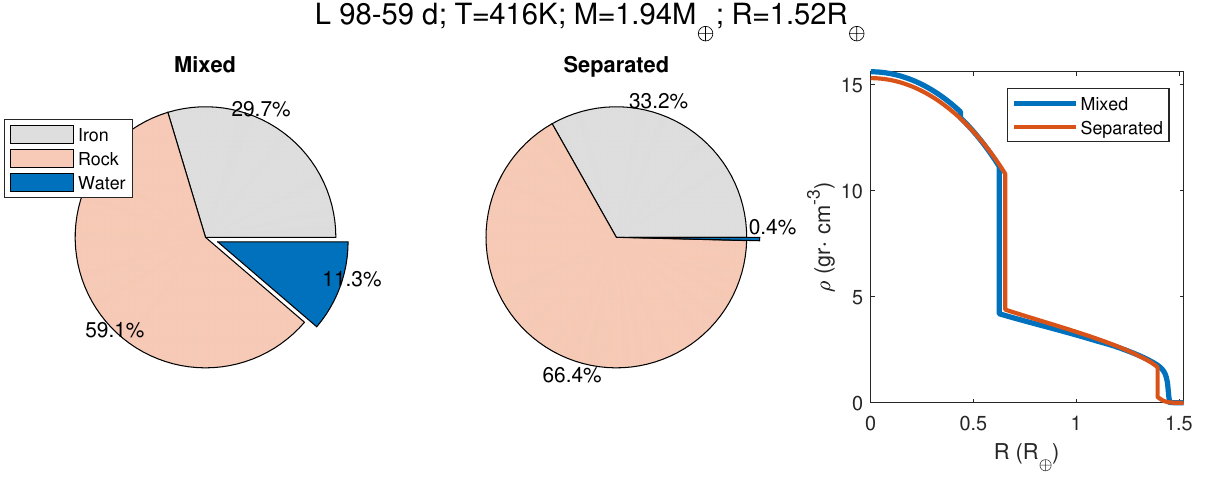}
    }
    \subfigure{
        \includegraphics[width=0.55\linewidth, trim={0.5cm 10.5cm 0.0cm 11.0cm},clip]{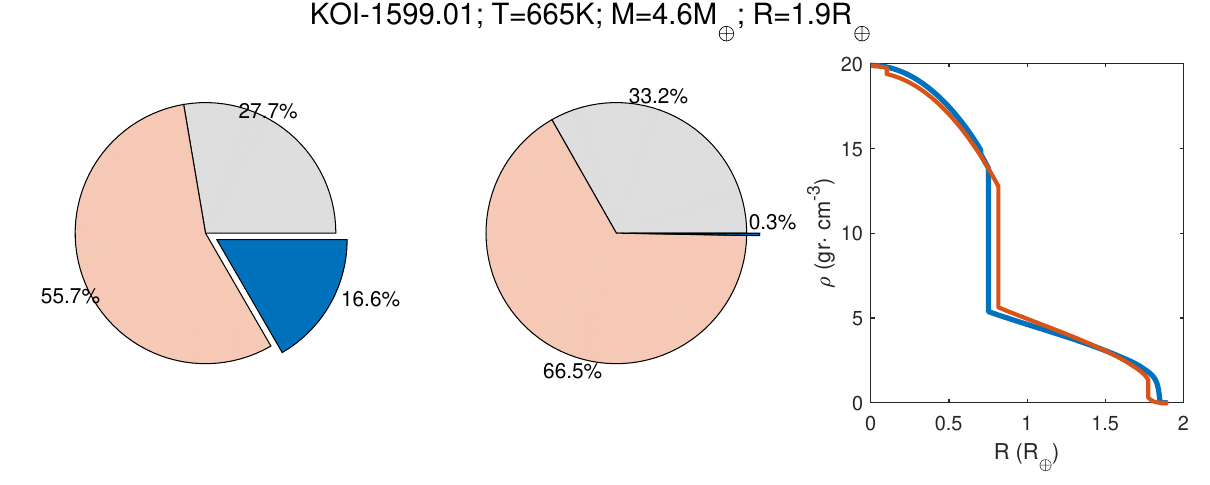}
    }
    \subfigure{
        \includegraphics[width=0.55\linewidth, trim={0.5cm 10.5cm 0.0cm 11.0cm},clip]{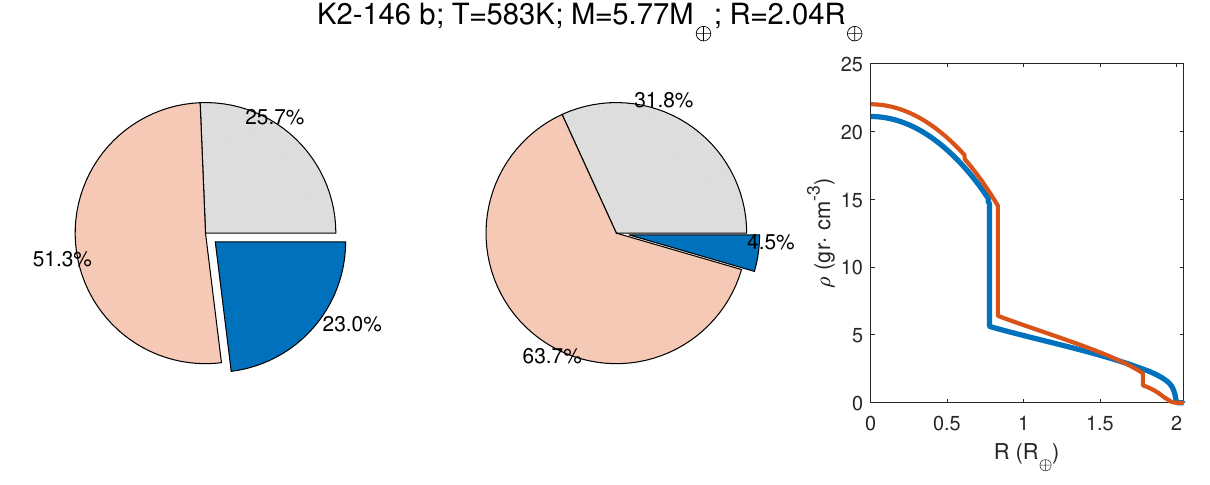}
    }
    \subfigure{
        \includegraphics[width=0.55\linewidth, trim={0.5cm 10.5cm 0.0cm 11.0cm},clip]{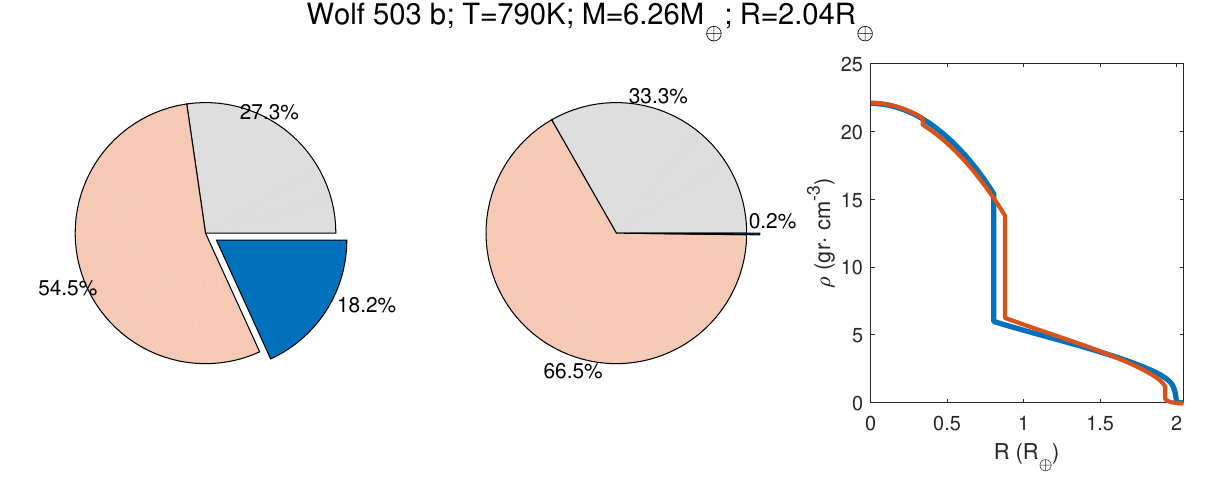}
    }
    \subfigure{
        \includegraphics[width=0.55\linewidth, trim={0.5cm 10.5cm 0.0cm 11.0cm},clip]{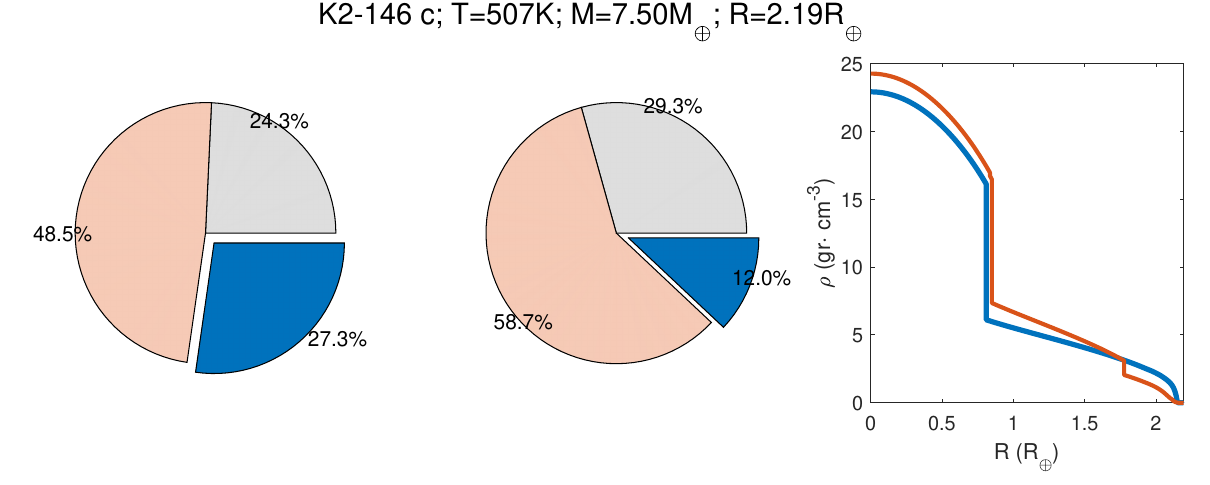}
    }
    \caption{{Internal structure of five exoplanets modeled in this study (Table \ref{Tbl:Pies}). The left pies show the bulk composition of planets with "mixed" models. The right pies show the bulk composition for "separated" models. The  right panels show density versus radius.} }
    \label{fig:Pies}
\end{figure}

\begin{table}[]
\centering
\begin{tabular}{ccccccccc}
\hline
\hline
Name               & $M$ ($M_{\oplus}$)        & $R$ ($R_{\oplus}$)       & $T_{\mathrm{eq}}$ (K) & WMF (Mixed) & WMF (Separated) & Reference                \\
\hline
L\,98-59 d         & $1.94 \pm 0.28$           & $1.52 \pm 0.12$          & $416 \pm 20$         & $0.113$    & $0.004$         & \citet{Demangeon2021}    \\
{KOI-1599.01}      & $4.6 \pm 0.3$             & $1.9 \pm 0.3$            & $665$                & $0.166$         & $0.003$             & \citet{Panichi2019}      \\
K2-146 b           & $5.77 \pm 0.18$           & $2.04 \pm 0.06$          & $583$                & $0.230$    & $0.045$         & \citet{Hamann2019}       \\
Wolf 503 b        & $6.26^{+0.69}_{-0.70}$    & $2.04 \pm 0.07$          & $790 \pm 15$                  & $0.182$         & $0.002$             & \citet{Polanski2021}     \\
K2-146 c           & $7.50 \pm 0.23$           & $2.19 \pm 0.07$          & $507$                & $0.273$    & $0.120$        & \citet{Hamann2019}       \\
\hline
\end{tabular}
\caption{{Planets simulated in this study and their inferred water mass fractions: "separated" refers to a three-layer structure (iron, rock, and water) with sharp transitions between the layers; "mixed" refers to models with an iron core and a mixture of rock and water above, with the outermost shell being pure water (see section \ref{sec:mixture-separation}). References to the respective data sources of planetary properties ($M$, $R$, $T_{\text{eq}}$) are included. For the K2-146 system {and KOI-1599 system}, $T_{\text{eq}}$ was calculated in this study using Kepler's laws, {under assumption of zero albedo.}}}
\label{Tbl:Pies}
\end{table}

{In this section we are modeling a sample of observed exoplanets. Similarly to \citet{Luque2022} we assembled a sub-sample of exoplanets (Table \ref{Tbl:Pies}) that might be interpreted as “50\% water world”, based on \citet{Zeng2019}'s curve, but without restricting our sub-sample to particular equilibrium temperature or stellar type (Figure \ref{fig:MvrRexo}). We are using our methodology (see section \ref{sec:findWMF}) to calculate the WMF for each planet we selected. In contrast with \citet{Zeng2019}'s model, our structure models account for the combined effects of temperature, phase transitions, rock–water miscibility, and separation (see section section \ref{subsec:compos}). We find that under our models, water mass fractions span 0.002–0.273, rather than a uniform 0.5 that was assumed then selecting planets on a curve of “50\% water world” curve from \citet{Zeng2019}}.  
These variations {of WMF depend on planetary mass, radius}, equilibrium temperature, and model assumptions (such water-rock miscibility versus distinct layered structure).  The simulated interior structures are depicted in Figure \ref{fig:Pies} and summarized in table \ref{Tbl:Pies}. 
 {Figure \ref{fig:Pies} illustrates the bulk compositions of the examined exoplanets (see Table \ref{Tbl:Pies}), in both "separated" and "mixed" configurations. It should be emphasized that }in the "mixed" cases, the water is found in two regions: some of it is integrated within the rock, due to high pressures within the interior, while another portion exists separately from the rock at lower pressures, forming a distinct water layer near the surface. As the density profiles indicate, the water in the outer shell is in vapor {and super-critical phases. Figure \ref{fig:Pies} shows that} the temperature has very large influence on the inferred water content, where hotter planets are typically found to have less water. {Figure \ref{fig:Pies} can be compared in some level with with Figure \ref{fig:Hist5Me}: Figure \ref{fig:Pies} shows much stronger differences of WMF between "mixed" and "separated" cases for a given planetary radius compared to differences in radius between "mixed" and "separated" in Figure \ref{fig:Hist5Me} for a given WMF. This is a result of of the radius being a much stronger constraint on possible compositions than the WMF: Radius effectively narrows the range of potential WMF values, whereas WMF does not as strictly restrict the possible range of planetary radii. 

{It is interesting to note that some planets that were selected as WMF=0.5, based on \citet{Zeng2019}, are found to have less than WMF=0.01 in our "Separated" scenario. This result is explained by very different model assumptions, resulting in different M-R relations, as discussed in section  \ref{sec:MRothers} .}


\subsection{M-R Relations}
\label{sec:MR_rel}
In this section, we explore M-R relation as a function  of WMF and thermodynamic scenario ("mixed" or "separated"). The results are presented in Figure \ref{fig:MRTWMF300_500}. For low WMF {(typically below 0.05; depending on pressure and temperature)}, the SCP lies beneath the entire water layer, resulting in complete separation of water and rock. {Consequently, the "mixed" scenario cannot be used to model planets with very low WMF.}

The Figure indicates that separated models yield a a higher radius than mixed, and that the difference between mixed and separated models typically decreases with higher planetary mass and also typically decreases with higher WMF \footnote{ {for WMF sufficient to properly model a "mixed"-scenario planet} }. The case of a dry planet (WMF = 0) is corresponding to a model of pure rock mantle above rocky core with mass ration of 2:1.


{Figure \ref{fig:upper_part} shows the  density profile of a planet versus mass and radius. On the left panel, one can see the effect of equilibrium temperature on the radius, as the total radius increases with surface temperature. The right panel shows that as $T_{eq}$ increases, less water separates from the rock, since the SCP shifts closer to the surface. The figure illustrates the competing influences of thermal expansion and the reduction of the water layer with temperature. As $T_{eq}$ rises from 400 K to 700 K, the mass of pure water layer falls from \(1.02\times10^{-3}\,M_\oplus\) to \(1.06\times10^{-4}\,M_\oplus\) (right panel), while its layer width decreases only from \(4.78\times10^{-2}\,R_\oplus\) to \(4.33\times10^{-2}\,R_\oplus\) (left panel). Thus, an order-of-magnitude decline in water mass corresponds to merely a 10\% change in layer thickness. Consequently, the net increase in planetary radius with temperature is dominated by thermal expansion of the rock–water mantle, as the mass of the pure water layer drops sharply with  higher \(T_{\rm eq}\).}

\begin{figure}[h!]
    \centering
 \includegraphics[width=0.9\textwidth, trim={1.2cm 5.0cm 1.5cm 5.0cm},clip]{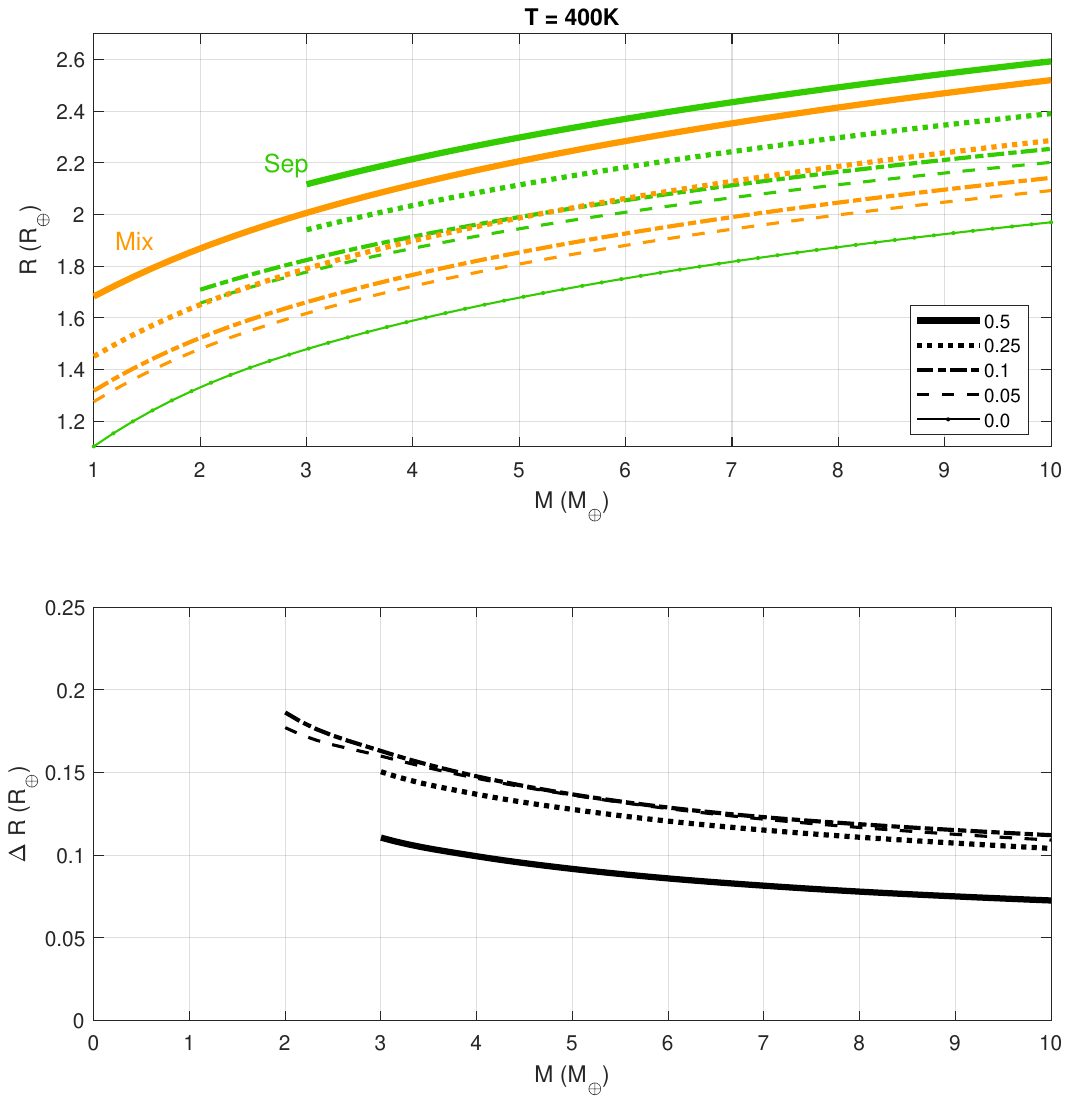}
        \caption{{M-R-WMF relations for "separated" (green) and "mixed"  (golden) models with $T_{eq}=400K$. The WMF is presented as a line style. "0.0" correspond to a model of a rocky planet with iron core, without water. The lower panels show the differences between separated and mixed radii for each WMF. }}\label{fig:MRTWMF300_500}
\end{figure}

\begin{figure}[h!]
    \centering
    \includegraphics[width=0.9\textwidth, trim={1.40cm 0.5cm 0.5cm 0.5cm},clip]{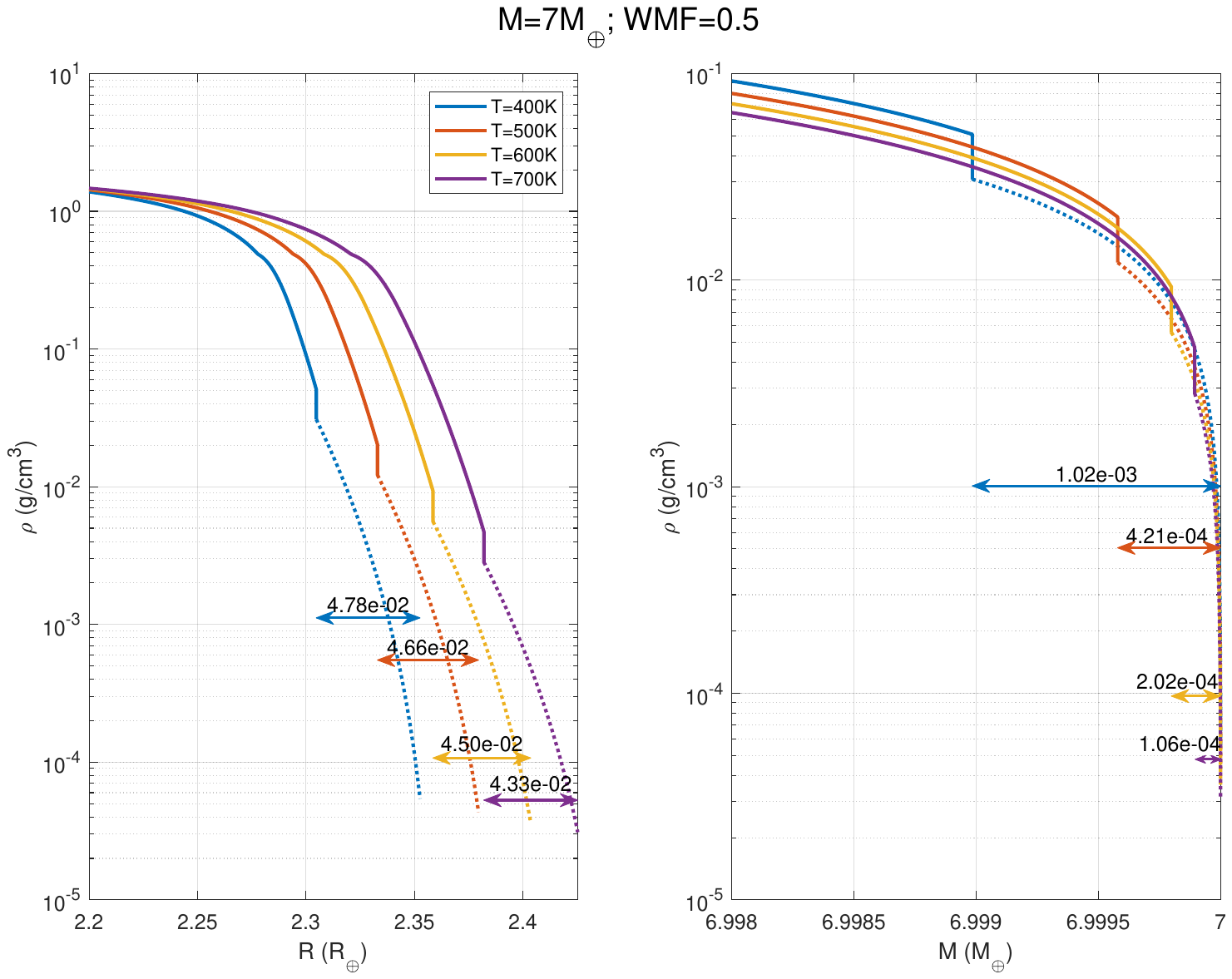}
        \caption{{Density versus radius (left) and mass (right) in the outermost layers of a 7$M_\oplus$ planet with WMF=0.50, under various temperatures in the "mixed" scenario. Solid lines represent water–rock mixtures; dotted lines indicate pure water regions. Double-headed arrows mark the extent of the water region in terms of radius or mass in Earth units. }}\label{fig:upper_part}
\end{figure}

\section{Discussion}
\label{sec:Discussion}

Our calculated planetary radii closely match the expected radii regions for water-rich super-Earths, as noted in previous studies \citep[e.g.][]{Lozovsky2018,Piaulet2023}. However, interior models do not provide a unique solution, as determining the water-to-rock ratio of these planets through observations is not possible, given that different internal models produce degenerate mass-radius-temperature profiles.

\subsection{Limitations}
\label{src:limits}
Several important considerations and limitations of this study should be acknowledged.

\begin{itemize}
    \item Our analysis does not account for gradual structural composition within the planets, instead we assume sharp boundaries {between regions} and homogeneous structure of each region. In this approach we investigate the two end cases (full separation and homogeneous mix), while the realistic case may lay somewhere between. Also, SCP is treated as a sharp transition between EOS's, where all the material below certain pressure and temperature is considered to be pure water, while above this point it is considered to be homogeneous mix. In reality this transition between mixed and separated  regions is gradual as well, creating {shallow transition zone below the pure water (this transition area mass is on the order of 1e-5 $M_\oplus$ for planets of orders of magnitude of 1 M$_\oplus$). Moreover, the miscibility itself is depending not only on P-T, but also on WMF itself \citep{Vazan2022,Dorn2021}, which is not taken into account. }
    \item While we have a comprehensive understanding of the EOS and phase transitions for pure rock and water, determining the EOS for the mixture of water and rock remains elusive. In this study, we approximated this EOS by taking a weighted average of the EOS for water and rock, utilizing an additive volume law. This method, as shown by \citet{Pan2023}, provides a reasonable approximation for water-rock mixtures.
    \item{ When modeling exoplanets, the uncertainties in the measured planetary masses, temperatures, and radii are not taken into account {as we used the central value only. Varying these parameters is expected to broaden} the values we find here. 
    \item The possible greenhouse effect of the atmospheric vapor is not included, although water is expected to trap-in some of the energy. {The trapped energy might change the temperature gradient and possibly extend the atmosphere \citep[e.g.][]{Boer2025}.}}
     \item {Larger planets, such as 5-10 M$_\oplus$ mass planets in relatively low temperatures ({about} 300K-400K) are expected to have significant percentage of H+He in their atmospheres (e.g. \citet {Madhusudhan2021,Rogers2023} and figure \ref{fig:MvRAugichne3} {in the Appendix}). } 
     \item  The rock EOS is taken to be FEOS \citep{Faik2018}, as we assume fused silica (SiO$_2$). However, our SCP is taken form \citet{Melekhova2007}, that was calculated for mixture or different rocks  {(MgO, SiO$_2$)} and not pure SiO$_2$ (see Appendix 1 table for details). 
    \item {Under high pressure and temperature, water can dissolve into an iron core\citep[e.g.][]{Schlichting2022,luo2024}. However, we focus on water-rich planets with an outer water shell and do not consider this effect in our models. Water dissolved in the core would have a minimal impact on the planetary radius compared to large reservoir of pure water on the surface, because the density of water under core T and P conditions is two orders of magnitude higher: For a planet of  5$M_\oplus$ planet at $T_\mathrm{eq} = 500$ (for example) the density of water at T-P of mantle-core-boundary is 4.28 g cm$^{-3}$ , compared to the surface water density of 0.02 g cm$^{-3}$. Consequently, the water in the outer shell {and in the mantle} dominates the contribution to the total radius. Therefore, including water-iron mixing is secondary to the effect on M-R relations {for planets with high WMF}. }
   \item   {In this work, we adopt a rock-to-iron mass ratio of 2:1, roughly reflecting Earth's composition, without considering volatiles mixed within the iron core (see previous point). As pointed by \citet{Huang2022} this ratio might underestimate 1 $M_\oplus$ planet radius by approximately 5\%, compared to Earth. While our chosen ratio may introduce some underestimation of the radius, it provides a reasonable approximation for the purposes of this study.}
    \item{{In high T and P regime, the AQUA EOS \citep{Haldemann2020} uses S(P,T) from an old version of \citet{Mazevet2019} to compute thermal gradients, in which the entropy  S(P,T)  was reported to be inaccurate. As a result, the adiabatic profiles derived from AQUA may not be fully reliable in high-temperature, high-pressure regimes, even though the P–T–$\rho$ relations remain consistent. Notably, the EOS by \citet{Mazevet2019} provides a good match to data from their \textit{ab initio} simulations, though some discrepancies remain when compared to experimental results. }} 
    \item {{Gradient calculations near cell boundaries in the AQUA EOS suffer reduced accuracy due to interpolation-induced discontinuities between adjacent cells (see section \ref{sec:EOSmodel}). To estimate the impact of this numerical effect, we compared our results with \citet{Haldemann2020}’s adiabatic pure-water planet model (5 \(M_\oplus\), \(T_{\text{eq}} = 500\) K). Our methodology reproduces their mass-radius results with a maximum radius discrepancy of 1.8\% in the worst case, which we attribute to the accumulated errors from numerical derivatives across table cells, as opposed to their smoother gradient computation.}}
\end{itemize}


These complexities underscore the challenges inherent in accurately characterizing the internal composition of exoplanets. Further research is needed to refine our understanding of their structural properties and to address these uncertainties effectively.

}

\subsection{Steam Atmosphere}
\label{sec:steam}

{{In this study we treat planets with atmospheres of pure water steam}. H+He was not implemented, as we aim to isolate the effect of water mass distribution on the radius, and H+He will obscure the main effect. Modeling planets with pure water atmospheres is essential for understanding the behavior of this specific substance and its influence on planetary structure and M-R relationships. However, it is crucial to keep in mind that such models represent highly idealized end-member scenarios and are not entirely realistic.}

As we demonstrate in this work, planets with high water content develop an outer water vapor shell at high $T_{eq}$, even if most of the water is contained within the rocky mantle. Hot water-rich exoplanets may encounter some thermal atmospheric escape, due to irradiation from their host stars, owing to their short orbital periods. One might speculate that this vapor shell simulated in this study could be subject to atmospheric loss. {This issue is addressed in the Appendix, section \ref{sec:ThermalEscape}, that shows that water atmosphere is unaffected by this phenomena.}

\subsection{M-R comparison to other works}
\label{sec:MRothers}

{Our results are comparable to other works in the field, as various groups have constructed planetary models with similar compositions. However, it is important to note that different models are based on distinct assumptions, leading to significantly different planetary structures and thus different M-R-T relationships.}

\begin{figure}[h]
    \centering
    \includegraphics[width=1.0\textwidth, trim={3.5cm 8.7cm 4.0cm 8.5cm},clip]{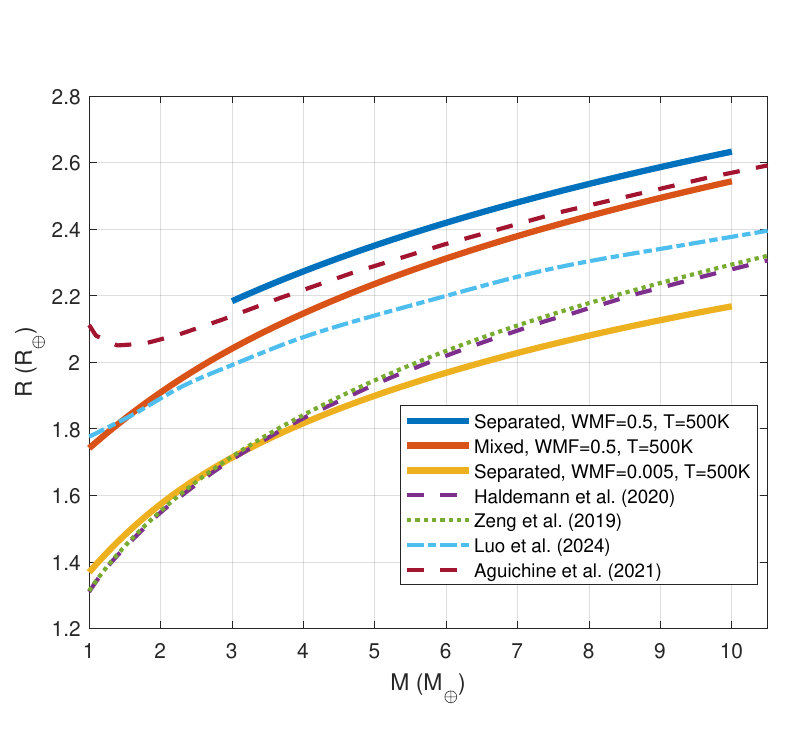}
    \caption{M-R relations from different works, compared with results of our models: the solid blue and red lines are "separated" and "mixed" for 0.5 water mass fraction and $T_{eq}=500K$. The yellow line is showing separated model for WMF=0.005,  $T_{eq}=500K$. The different models in this comparison have different assumptions, as listed in Table \ref{tab:comprMRw}}. 
    \label{fig:comprMRworks}
\end{figure}


Figure \ref{fig:comprMRworks} compares our results to those of previous studies. {The solid lines represent the cases of "separated" and "mixed" models for a planet with WMF=0.5 or 0.005, a surface pressure $P_{\text{surface}}$ of 100 mbar, and $T_{\text{eq}}=500\,\mathrm{K}$.} The assumptions and parameters for each compared work are summarized in Table \ref{tab:comprMRw}. 
{The differences in the M-R relations presented across the studies result from varying assumptions, such as chemical composition, surface pressure, equilibrium temperature, and material distribution.} {As this study used FOES (SiO$_2$) as the EOS of rock, our radii are typically higher than other works in this comparison.}

{It is important to emphasize that our "separated" M–R relations differ substantially from those presented by \citet{Zeng2019}, which were used to select a sub-sample for Section~\ref{sec:exopl}. As a result, the inferred WMF's of exoplanets (Figure \ref{fig:Pies})  differ by orders of magnitude from the cited value of 0.5. A key distinction lies in the assumed thermal structure: \citet{Zeng2019} (and later \citet{Haldemann2020}) adopted an isothermal atmosphere, which leads to an underestimation of the planetary radius: in such a model, pressure increases with depth without any increase in temperature, thereby underestimating the thermal expansion of the gas. In contrast, we demonstrate that the same observed radius can be reproduced with significantly less water when an adiabatic thermal profile is assumed. }

\begin{table}[h!]
\setlength{\tabcolsep}{1pt} 
\renewcommand{\arraystretch}{1.1} 
\centering
\begin{tabular}{|l|l|l|l|l|l|}
\hline
{Paper} & {Surface} & {Bulk Composition} & {Metal} & {Rock} & {Notes} \\ \hline
This study (Mixed/Separated) & $T_{\text{eq}} = 500$ K & WMF=0.5 & Fe & SiO$_2$ & \\ 
                             & $P = 100$ mbar & Iron:Rock=1:2 &  &  & \\ \hline
\citet{Haldemann2020}  & $T_{\text{eq}} = 500$ K & WMF=0.5 & Fe & MgSiO$_3$ & Isothermal structure \\ 
                       & $P = 1$ mbar & Metal:Rock $\approx$ 1:2 &  &  & \\ \hline
\citet{Zeng2019}       & $T_{\text{eq}} = 500$ K & WMF=0.5 & Fe, Ni & MgSiO$_3$ & Isothermal structure;\footnote{\url{https://lweb.cfa.harvard.edu/~lzeng/planetmodels.html\#planetinteriors}}  \\ 
                       & $P = 1$ mbar & Metal:Rock $\approx$ 1:2 &  &  &  \\ \hline
\citet{luo2024}        & $T_{\text{eq}} = 400$ K & WMF=0.5 & Fe & MgO, SiO$_2$, FeO, & Includes water dissolution \\ 
                       & $P = 1$ mbar & Iron:Rock=1:2 &  & (Mg,Fe)SiO$_3$, Mg$_2$SiO$_4$ & in mantle and core \\ \hline
\citet{Aguichine2021}  & $T_{\text{eq}} = 500$ K & WMF=0.5 & Fe, FeS & (Fe,Mg)SiO$_3$, (Fe,Mg)O, & Includes effects of irradiation \\ 
                       & $P = 10^{-3}$ mbar & Metal:Rock $\approx$ 1:2.3 &  & (Fe$_2$,Mg$_2$)SiO$_4$, (Fe$_2$,Mg$_2$)Si$_2$O$_6$ & and evaporation\footnote{\url{https://archive.lam.fr/GSP/MSEI/IOPmodel/mr_all.dat}} \\ \hline
\end{tabular}
\caption{{Summary of properties in works similar to this study, that are shown in Figure \ref{fig:comprMRworks}; $T_{\text{eq}}$ (equilibrium temperature) and $P$ (surface pressure) represent boundary conditions in these planetary models. "Metal" is the chemical composition of the metallic core, and "Rock" is the chemical composition of the rocky mantle. }}
\label{tab:comprMRw}
\end{table}

\section{Summary and conclusions}

This study investigates the internal structure of water-rich sub-Neptune planets, focusing on the distribution and state of water within these planets. By utilizing MAGRATHEA internal structure code, we modeled planets with masses ranging {up to} 10 M$_\oplus$ and varying water mass fractions, temperatures and examined scenarios with and without considering water-rock miscibility. These two scenarios actually bracket the possible rock and water distributions. 

In in previous works \citep{Dorn2021,Kite2021,Vazan2022} it has been shown that water can be dissolved in the rocky mantle of planets. We use it as a foundational aspect of our analysis. Considering the miscibility of water in the mantle is crucial to infer water mass fraction for a planet, and therefore for construction of M-R relations.

This work addresses two key opposing effects: firstly, when water is mixed and dissolved in the rocky mantle, the planetary radius decreases; secondly, at high temperatures, {both the mantle and water vapor that is} forming an outer shell {expand}, causing the radius to increase. By accounting for these effects, we highlight significant uncertainties and differences in planetary radius resulting from the distribution of water within the planet's interior. Both water miscibility within the mantle and the behavior of the outer water layer, strongly influenced by temperature, play important roles in these variations.

 
The main points can be concluded as:

\begin{enumerate}
      \item { Impact of water distribution}: Considering the water-rock miscibility in determining the interior structure has a substantial impact on the planets' radii. Fully "separated" models generally exhibit larger radii for a given mass, equilibrium temperature, and composition than "mixed" models, especially at higher temperatures. This difference arises because the homogeneous rock-water mixture is denser than a pure water layer, leading to more compact planetary structures in "mixed" models.
    \item {Impact of temperature on structure}: For water-rich planets with temperatures and pressures below the SCP in their outer regions, a shell of pure water forms. At higher equilibrium temperatures, a water can sublimate and steam atmosphere can develop. Comparing "mixed" and "separated" scenarios, higher equilibrium temperatures result in more pronounced differences in the planets' radii between the two scenarios.

      \item {Increase in equilibrium temperature leads to smaller mass of the water layer, as larger fraction of water is mixed in the rocky mantle. However, the total radius of the planet is increasing due to thermal expansion.}
    \item  {Modeling exoplanets:} Comparing our models with observed exoplanets {that could be} indicated as having 50\% water, we found that the actual water mass fraction is significantly smaller (see Table \ref{Tbl:Pies}). The water mass fraction was calculated for each planet based on its observed mass, radius, equilibrium temperature, and two assumed internal structure end-members. 

\end{enumerate}


In conclusion, our investigation into the internal structures of water-rich exoplanets reveals the critical role of pressure-temperature conditions in determining the distribution of water and rock. These findings have significant implications for interpreting observational data and advancing our understanding of these celestial bodies. Future research should focus on refining the equations of state for rock-water mixtures and exploring the gradual transitions between mixed and separated states to enhance the accuracy of planetary models. Additionally, incorporating detailed atmospheric loss processes could provide further insights into the long-term evolution of water-rich exoplanets.

\section{Acknowledgments}
{We thank the referee for valuable comments which significantly improved our paper.} The illustration (figure \ref{fig:Illustration}) for this paper was done using publicly available latent diffusion model for text-to-image synthesis, SDXL \citep{Rombach2022,Podell2023}. Allona Vazan acknowledges support from the Israel Science Foundation (ISF) grants 770/21 and 773/21. {We thank David Rice for his assistance with the MAGRATHEA code and for providing several test cases that helped calibrate our models.{ M.L. thanks Allona Vazan for her support and important contributions to this study.}}

\bibliography{referenses}{}
\bibliographystyle{aasjournal}

\section{Appendix}
\subsection{Second Critical Point}
\begin{figure}[h!]
    \centering
    \includegraphics[width=0.8\textwidth, trim={2.0cm 6.7cm 2.4cm 7.5cm},clip]{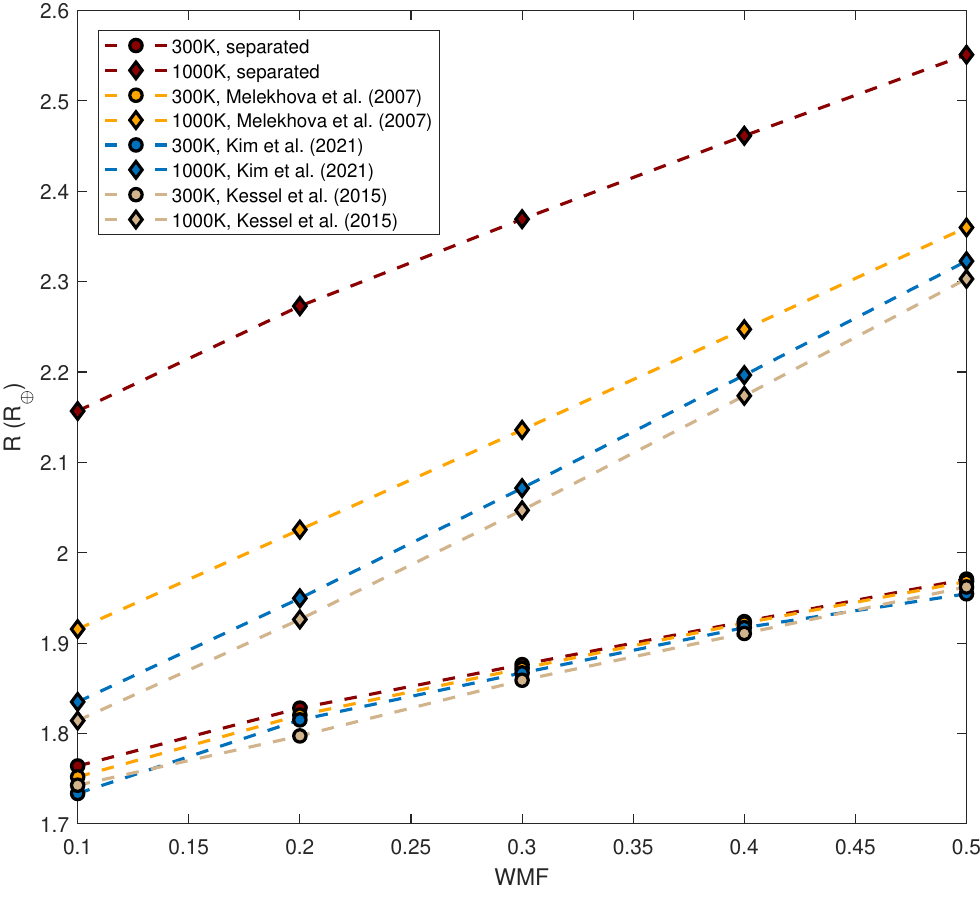}
        \caption{{Planetary radius versus water mass fraction (WMF) for planets assuming full water and rock separation (red lines), and planets where the separation occurs in the upper part according to different pressure and temperature conditions (SCP). {Planetary mass is chosen to be 5$M_\oplus$} and rock-to-iron ratio is 2:1 for all the cases. The circles correspond to models with equilibrium temperature of 300K, while diamonds corresponds to an equilibrium temperature of 1000K. Yellow, blue and tan colored lines examine different SCP's criteria for different solids as listed in Table \ref{TableSCP}. {Note that 300K models are pre-runaway greenhouse planets \citep[e.g.][]{Kopparapu2013,Turbet2019} and not shown in the main study.}  }}
    \label{fig:RvcPer}
\end{figure}

In this study, we utilized the critical conditions for water to separate from the rock-water mixture (SCP), as described by log(T) = 3.3 (in K) and log(P) = 1.1 (in GPa) following the methodology outlined by \citet{Melekhova2007}. However, an alternative criteria are also available (See table \ref{TableSCP}). This variance resulted in some differences in the calculated radii, as illustrated in Figure \ref{fig:RvcPer}. In all the cases, the "separated" cases have larger radii then every "mixed" scenario. There are differences in the calculated radii for different mixed cases, with \citet{Kim2021} giving the highest radius and \citet{Kessel2015} giving the lowest.  It is crucial to note that in this section, we solely examined the isolated influence of SCP, without altering the EOS or rock, which is based on \citet{Faik2018} {for all the cases}. It should be noted as well that other approaches to miscibility of rock and water can be applied \citep[e.g.][]{Kovacevic2022,Kovacevic2023}. {The models with $T_{eq}=300 K$ that presented in this section are pre-runaway models, in contrast with the post-runaway models presented in the main part.}


\begin{table}[h]
\centering
\begin{tabular}{|l|l|l|l|}
\hline
paper                   & log(T), K & log(P), GPa & composition                                \\ \hline
Melekhova et al. (2007) & 3.3       & 1.1          & $\text{MgO, SiO}_2, \text{H}_2\text{O}$                                   \\ \hline
Kessel et al. (2015)    & 3.1       & 0.7          & $\begin{array}{@{}l@{}}\text{SiO}_2, \text{Al}_2\text{O}_3, \text{Cr}_2\text{O}_3, \text{TiO}_2, \\ \text{FeO}, \text{MgO}, \text{CaO}, \text{Na}_2\text{O}, \text{K}_2\text{O}, \text{H}_2\text{O}\end{array}$ \\ \hline
Kim et al. (2021)       & 3.2       & 1.5          & $\text{MgO}$                                        \\ \hline
\end{tabular}
\caption{Second critical point of rock-water mixture, according to different studies.}
\label{TableSCP}
\end{table}

\subsection{Entropy Assumptions}

{Then developing Equation \ref{dTdPeq2} we assume that the entropy of a water-rock mixture can be expressed as $S = X_w S_w + X_r S_r$,
where \( X_w + X_r = 1 \) represent the mass fractions of water and rock in the mixture, respectively. This assumption implies ideal mixing, that the entropy of mixing is negligible compared to the entropies of water (\( S_w \)) and rock (\( S_r \)) combined, and that the entropies of the mixture's components are independent of one another. One might see the Appendix of \citet{Vazan2013} for details and assumptions of an ideal mixing. 
 }

{To find the temperature derivative with respect to pressure at constant entropy for the mixture, we applying the chain rule and using Maxwell relations to get:}
\begin{equation}
\left( \frac{dT}{dP} \right)_{S} = - \frac{\left( \frac{\partial S}{\partial P} \right)_{T}}{\left( \frac{\partial S}{\partial T} \right)_{P}},\label{dTdpEq}
\end{equation}
{
We are substituting the expression for \( S \) of mixture into the partial derivative in the numerator of Equation \ref{dTdpEq}:}
\[
\left( \frac{\partial S}{\partial P} \right)_{T} = \frac{\partial}{\partial P} \left( X_w S_w + X_r S_r \right)_{T} = X_w \left( \frac{\partial S_w}{\partial P} \right)_{T} + X_r \left( \frac{\partial S_r}{\partial P} \right)_{T}.
\]
{We apply the same to the denominator of equation \ref{dTdpEq}:}
\[
\left( \frac{\partial S}{\partial T} \right)_{P} = X_w \left( \frac{\partial S_w}{\partial T} \right)_{P} + X_r \left( \frac{\partial S_r}{\partial T} \right)_{P}.
\]

{Substituting these two expressions back to Equation \ref{dTdpEq}, we obtain:}
\[
\left( \frac{dT}{dP} \right)_{S, \text{mix}} = -\frac{X_w \left( \frac{\partial S_w}{\partial P} \right)_{T} + X_r \left( \frac{\partial S_r}{\partial P} \right)_{T}}{X_w \left( \frac{\partial S_w}{\partial T} \right)_{P} + X_r \left( \frac{\partial S_r}{\partial T} \right)_{P}}.
\]
{This result gives the temperature-pressure dependence at constant entropy for the water-rock mixture as, from Equation \ref{dTdPeq2}.}

\subsection{{Thermal Escape}}
\label{sec:ThermalEscape}

\begin{figure}[h!]
    \centering
    \includegraphics[width=0.8\textwidth, trim={1.7cm 6.5cm 2.6cm 6.8cm},clip]{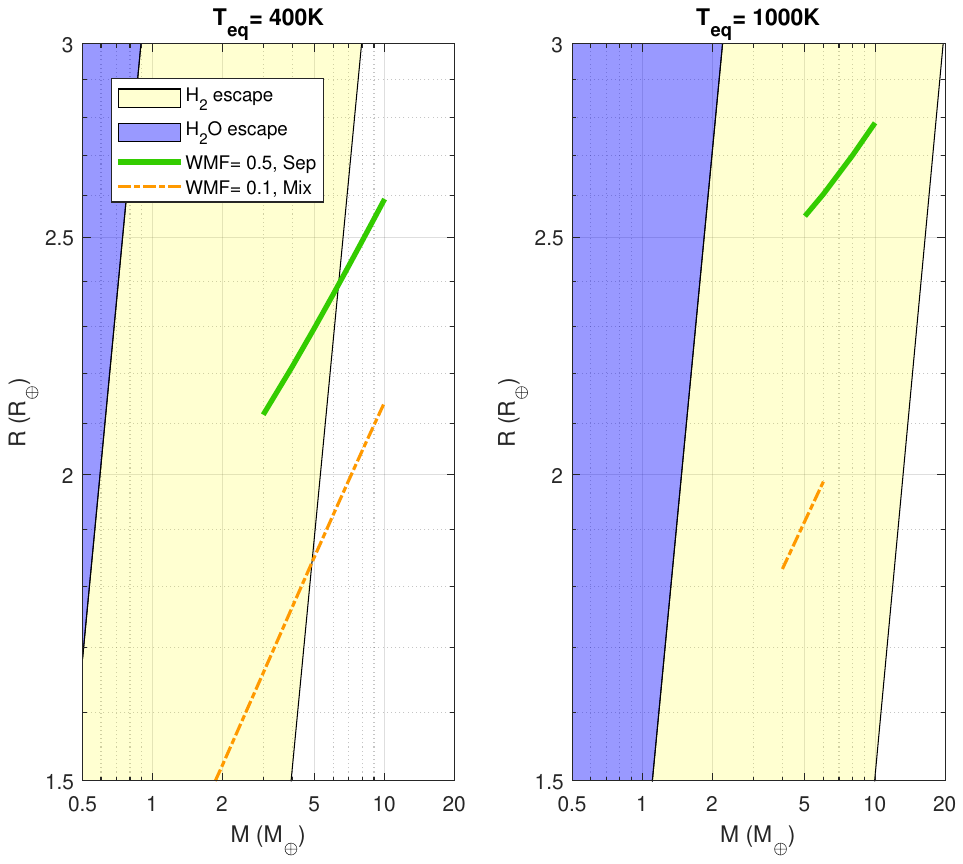}
        \caption{{{Mass–Radius diagram illustrating regions of $H_2$ and $H_2O$ atmospheric thermal escape (shaded areas) alongside modeled planetary M–R relation curves at equilibrium temperatures of 400 K (right) and 1000 K (left). The curves correspond to water mass fractions of 0.1 (separated scenario) and 0.5 (mixed scenario). The shaded areas are taken from \citet{Aguichine2021}.} }}
    \label{fig:MvRAugichne3}
\end{figure}

{One can claim that that water-steam atmosphere (section \ref{sec:steam}) might the subject to mass lose via evaporation . However, as depicted in Figure \ref{fig:MvRAugichne3} , our simulated planets reside in mass-radius-temperature regime significantly below the water escape threshold. This finding aligns with \citet{Aguichine2021} and \citet{Kurosaki2023}, who determined that water is much less prone to atmospheric escape than H+He. We should note {again} that in this study, as we focus on the effect of the water distribution on the radius, the H+He in the atmosphere is not studied, although in highers mass and lower temperature a planet might retain some fraction of light gases.}

\end{document}